# Growth mechanism of superconducting $MgB_2$ films prepared by various methods

H.Y. Zhai, H.M. Christen, L. Zhang, M. Paranthaman, C. Cantoni, B.C. Sales, P.H. Fleming, D.K. Christen, and D.H. Lowndes

Oak Ridge National Laboratory, Oak Ridge, TN 37931-6056

The growth mechanisms of $MgB_2$ films obtained by different methods on various substrates are compared via a detailed cross-sectional scanning electron microscopy (SEM) study. The analyzed films include (a) samples obtained by an *ex-situ* post-anneal at 900°C of e-beam evaporated boron in the presence of an Mg vapor (exhibiting bulk-like $T_{c0}$ ~ 38.8 K), (b) samples obtained by the same *ex-situ* 900°C anneal of pulsed laser deposition (PLD)-grown Mg+B precursors (exhibiting $T_{c0}$ ~ 25 K), and (c) films obtained by a low-temperature (600 – 630°C) *in-situ* anneal of PLD-grown Mg+B precursors (with $T_{c0}$ ~ 24 K). A significant oxygen contamination was also present in films obtained from a PLD-grown precursors. On the other hand, it is clearly observed that the films obtained by the high-temperature reaction of e-beam evaporated B with Mg vapor are formed by the nucleation of independent $MgB_2$ grains at the film surface, indicating that this approach may not be suitable to obtain smooth and (possibly) epitaxial films.

The discovery of superconductivity at 39K in the binary inter-metallic compound magnesium diboride ($MgB_2$) *(1,2)* has spurred intense research efforts worldwide. Hall measurement confirmed holes as the majority carriers *(3)* and isotope effect studies for both magnesium and boron indicated classic BCS behavior *(4,5)*. However, there is evidence of two-dimensional superconductivity on the B layer with the Mg layer contributing to the carrier density *(6)*, and an upper critical field anisotropy ($H_{c2}^{ab}/H_{c2}^{c} \approx 1.7$) has been observed *(7)*. These properties distinguish $MgB_2$ from isotropic BCS superconductors, and an understanding of its behavior may shed light onto issues related to high-$T_c$ cuprates. Unfortunately, crystal growth of $MgB_2$ has proven elusive, mainly due to the stability of B-rich compounds ($MgB_4$, $MgB_7$) that form at higher temperature *(8)*, and sintered $MgB_2$ ceramics rarely reach densities above 80%. For future experimental work, dense, highly textured or epitaxial thin films are desirable.

A significant experimental effort has been focused onto the growth of $MgB_2$ films. To the best of the authors' knowledge, all approaches described to date rely on a process consisting of film growth at low temperature, followed by a high-temperature anneal. In fact, formation of the correct stoichiometry by depositing directly at temperatures above 300°C has proven elusive *(9)*. The highest transition temperatures are observed in films obtained by a method in which an elemental boron film is reacted at elevated temperature (≈ 900°C) in a Mg-rich environment *(10-13)*, a process similar to that first applied to the fabrication of $MgB_2$ wires *(14)*. Films obtained via an *in-situ* anneal of Mg+B precursors at relatively low temperature (≈ 600°C) typically exhibit $T_c \approx 20 – 25$ K *(10, 13, 15-17)* and poor crystallinity, but *in-situ* approaches may be necessary to fabricate junctions, and ultimately to obtain epitaxial films.

The present study aims to elucidate the different mechanisms for nucleation and growth of $MgB_2$ in the various previously described routes and will illustrate the striking differences between these approaches.

Three types of samples are prepared and analyzed in this work:

1) "B + *ex-situ*": E-beam evaporated boron, reacted at 900°C in a crimped Ta cylinder containing an excess amount of $MgB_2$ and elemental Mg, as described in *(10)*. Sample A was annealed for 20 minutes, and sample B was annealed for 1 hour. These films are characterized by $T_{c0} \approx 38.8$ K with a transition width of ~ 0.2K.
2) "PLD + *ex-situ*": PLD-grown precursors (either $MgB_2$ deposited by ablation from a stoichiometric target, or Mg-rich $MgB_2$ deposited by ablation from a segmented target formed by one half of $MgB_2$ and one half of elemental Mg), subjected to the same annealing process as the "B + *ex-situ*" films. These films exhibit $T_{c0}$ ~ 25 and $T_c^{onset}$ ~ 28K when grown in a PLD system with a background pressure of $2 \times 10^{-4}$ Torr Ar/4%$H_2$.
3) "PLD + *in-situ*": PLD-grown precursors (Mg-rich $MgB_2$ deposited by ablation from a segmented target as above, or stoichiometric $MgB_2$), annealed *in-situ* in vacuum or Ar/4%$H_2$. These films, grown in the PLD system with a background pressure of $3 \times 10^{-6}$ Torr exhibit $T_{c0}$ ~ 24K with narrow transitions (≈ 1K).

Results from R(T) measurements performed on samples from each of these three groups are summarized in Table 1. Different substrates have been used for each group (including R-plane and C-plane $Al_2O_3$, $LaAlO_3$, $SrTiO_3$, and Si). The differences between films on dissimilar substrates within the same group are small as compared to the differences between the different groups. High-temperature anneals (group 1 and 2) were not successful with Si substrates due to a significant interdiffusion at the film/substrate interface.

For several of the samples listed in Table I, a detailed scanning electron microscopy (SEM) and energy dispersive spectroscopy (EDS) investigation was performed. Figure 1 shows the results for samples from the first category ("B + *ex-situ*"), i.e. obtained by reacting a 500nm thick e-beam evaporated B film on a $Al_2O_3$ substrate at 900°C for 20 minutes (sample A in Table. 1, Figure 1a,b) and 1 hour (sample B in Table. 1, Figure 1c,d). The surface shows well-defined $MgB_2$ crystallites, for which no evidence of oxygen contamination was found in EDS spectra. In these spectra, the concentration of B is difficult to determine quantitatively. Qualitatively, however, it was observed from data taken on the cross-section of a cleaved sample that the denser portion of the film had a stronger signature of oxygen than the surface region of the sample. A well-defined interface layer was present in the sample annealed for 20 minutes (Figure 1b). The sample annealed for 1 hour was dense without a visible interface layer, but tended to delaminate from the substrate.

Figure 2 shows oblique-view and cross-section SEM images for sample C from the second category ("PLD + *ex-situ*"), i.e. obtained from a $MgB_2$ precursor, and film deposited by PLD in a background $2\times10^{-4}$ Torr of Ar/4%$H_2$. The sample was then annealed *ex-situ* at 900°C for 1 hour in the presence of excess Mg (like sample B, annealing 1 hour). Porosity and non-uniform grain formation are clearly observed in the cross-section image, as is an interface layer. Contrary to the observation in Figure 1b, this interface structure actually consists of two distinguishable layers, the origin of which is unclear at this time.

Plan-view and cross-section SEM images for sample D are shown in Fig. 3. This sample (from the "PLD + *in-situ*" group) was obtained by reacting a Mg-rich Mg+B precursor, capped with a layer of Mg, at 600°C in 0.7 atm of Ar/4%$H_2$ (as described in *Ref. 9)*. As compared to Figures. 1 and 2, this film appears dense and uniform. Well-defined, "egg-shell"-like hollow spheres of MgO (as determined by EDS) are observed, as is a thin MgO "skin" covering portions of the sample (see Fig. 3b). The existence of the "egg-shell"-like structures can be explained by noting that the PLD-grown Mg films showed droplets of a comparable size (0.5 – 1.5 μm) as is often observed in PLD of metals. The formation of these well-defined MgO structures indicates that a significant amount of oxidation occurs before the magnesium reacts or evaporates. It is also important to note that the Mg target itself contained an amount of oxygen on the order of a percent *(9)*.

Finally, Fig. 4 shows an SEM image of sample E (“PLD + *in-situ*”), i.e. obtained by *in-situ* annealing of a PLD precursor in$10^{-4}$ Torr of Ar/4%$H_2$, at 630°C. The precursor for this sample was a simple stoichiometric layer of $MgB_2$ deposited at room temperature by PLD on Si, same as sample C (Figure 2). As shown previously *(15)*, the optimal annealing temperature of these “un-capped” $MgB_2$ precursors on Si was slightly higher than that for sample D described above. Sample E, characterized by $T_{c0}$ ~ 24 K, with a transition width of ~ 1 K, exhibits a clean surface and dense structure; however, the film easily delaminated from the substrate and showed signs of cracking. Similar cracking was observed on all samples of the “PLD + *in-situ*” group, regardless of the substrate used. While delamination was also observed in group 1 (Figure 1d), the delamination was most pronounced on the Si substrate.

EDS data suffer from a large experimental error on the B concentration. However, a qualitative comparison of the oxygen-to-magnesium ratio can be made. As noted above, sample A and B (from the “B + *ex-situ*” group) show large grains containing no detectable amount of oxygen, while the portion of in the middle of the films clearly exhibits a small but detectable amount of oxygen. Samples C (“PLD + *ex-situ*”), D, and E (both “PLD + *in-situ*”) were also characterized in detail, and the films exhibited significantly larger amounts of oxygen than samples A and B. The origin of the oxygen may be two-fold. First, oxygen was detected in the PLD targets *(9)*, and is thus likely to get incorporated via the laser plume. Second, atomic or ionized magnesium in the laser plume is likely to react readily even with small amounts of oxygen or water in the PLD system, and due to the forward-oriented nature of the deposition process, most of the formed MgO will be incorporated in the film. The situation in the *ex-situ* reaction is quite different: the Ta cylinder acts as an oxygen getter, and as the ampoule containing both the boron film and excess magnesium is heated, only a relatively small fraction of the total evaporated Mg actually reacts to form $MgB_2$, while most of the material deposits on the walls of the ampoule. Thus the Mg vapor can act as a further oxygen getter without necessarily contaminating the film. Finally, reacting a B film (rather than a mixture containing Mg) has the advantage that oxidation at room temperature during transfer of the precursor film from the deposition system to the annealing furnace is likely to be insignificant, whereas a precursor film containing Mg will readily react to form MgO.

The most significant result of the cross-section SEM images (Figs. 1-4) can be drawn by comparing the sample grown in the "B + *ex-situ*" process with all other samples. Clearly samples A and B with the highest, bulk-like $T_c$ are formed in a process that resembles crystal growth much more than ordinary film growth. In fact, the crystals appearing at the surface of the sample clearly nucleate independently of each other. This renders this process a very unlikely candidate for epitaxial film growth.

On the other hand, samples grown from a precursor containing both Mg and B show a more uniform profile. Inhomogeneities are only observed after high-temperature *ex-situ* anneals (Figure. 2). The inclusions in that film (sample C) are very different in nature from the $MgB_2$ grains in samples A and B, but their origin and composition have not been determined. It is interesting to observe the negative correlation between crystallization and oxygen contamination. While we have insufficient data to identify cause and effect, it appears likely that contamination inhibits crystallization of large grains.

To summarize, the present SEM study of $MgB_2$ films obtained by various methods illustrates very different types of phase formation and crystallization for the various approaches. It is clearly seen that in an approach consisting of the reaction of a B film with Mg vapor, $MgB_2$ grains grow independently of each other and of the substrate, with the nucleation of these grains likely to occur at the film surface. This approach may thus be an unlikely candidate for obtaining smooth $MgB_2$ films and layers with a structural relationship to the substrate. However, the method results in comparatively oxygen-free, well-crystallized grains, and material with a bulk-like $T_c$. All currently studied approaches yielding Mg+B mixed precursors suffer from a chronic oxygen contamination, which appears to limit grain formation. However, those films are comparatively dense and uniform.

The authors would like to acknowledge D. P. Norton and C. Rouleau for technical assistance and useful discussions. This research is sponsored by the U.S. Department of Energy under contract DE-AC05-00OR22735 with the Oak Ridge National Laboratory, managed by UT-Battelle, LLC, and by the DOE Office of Energy Efficiency and Renewable Energy, Office of Power Technologies – Superconductivity Program.

Figure captions:

Figure 1. Plan-view and cross-section SEM images of sample A (see Table I) annealed ~ 20 minutes (a, b), and Sample B annealed for 1 hour (c, d). These $MgB_2$ films on $Al_2O_3$ were formed by an *ex-situ* reaction of an e-beam evaporated B film on $Al_2O_3$ at 900°C in the presence of a Mg vapor. Well-formed $MgB_2$ crystals with sizes up to 1 μm are observed. An interface layer (approximately 30 nm thick) is visible in the sample annealed fro 20 minutes (Figure 1b).

Figure 2. Oblique-view and cross-section SEM images of sample C (see Table I). This $MgB_2$ film was formed by reaction of a PLD-grown $MgB_2$ film on $Al_2O_3$ then *ex-situ* annealed at 900°C in the presence of a Mg vapor (same annealing process as sample B, annealing time ~ 1 hour). Some porosity and a double interface layer (inset) are observed.

Figure 3. Plan-view and cross-section SEM images of sample D (see Table I). This $MgB_2$ film on $Al_2O_3$ was formed by an *in-situ* anneal at 0.7 atm. of Ar/4%$H_2$ at 600°C. The PLD-grown precursor consisted of a Mg-rich Mg+B mixture, capped with a layer of Mg. The visible droplets are “egg-shell”-like structures (see inset) formed by the partial oxidation of Mg droplets and the subsequent evaporation or reaction of the metallic Mg. Also visible is a MgO “skin” covering the dense, uniform film.

Figure 4. Oblique-view and cross-section SEM images of sample E (see Table I). This $MgB_2$ film on Si was formed by an *in-situ* anneal at 630°C in $10^{-4}$ Torr of Ar/4%$H_2$ of a stoichiometric, PLD-grown precursor.

Table I. Characteristics of R(T) measurements obtained for different samples. The various "groups" are defined in the text *(9-11, 15)*.

| | Substrate | Group | $R_{300K}$ (Ω) | $R_{onset}$ (Ω) | $T_c^{onset}$ (K) | $T_{c0}$ (K) | Precursor | Anneal |
|---|---|---|---|---|---|---|---|---|
| A | $Al_2O_3$ | 1 | 0.36 | 0.15 | 38.3 | 38 | B (e-beam) | 900 °C 20 mins |
| B | $Al_2O_3$ | 1 | 0.22 | 0.08 | 39.0 | 38.8 | B (e-beam) | 900 °C 1 hour |
| C | $Al_2O_3$ | 2 | 25 | 28 | 28.6 | 25.2 | $MgB_2$ at $2\times10^{-4}$ $Ar/H_2$ (PLD) | 900 °C 1 hour |
| D | $Al_2O_3$ | 3 | 22.2 | 22 | 28.5 | 24.6 | $Mg/MgB_2$ at $2\times10^{-4}$ $Ar/H_2$ (PLD) | 600 °C 20 mins |
| E | Si | 3 | 0.8 | 21.4 | 25 | 24.2 | $MgB_2$ at $2\times10^{-4}$ $Ar/H_2$ (PLD) | 630 °C 20 mins |

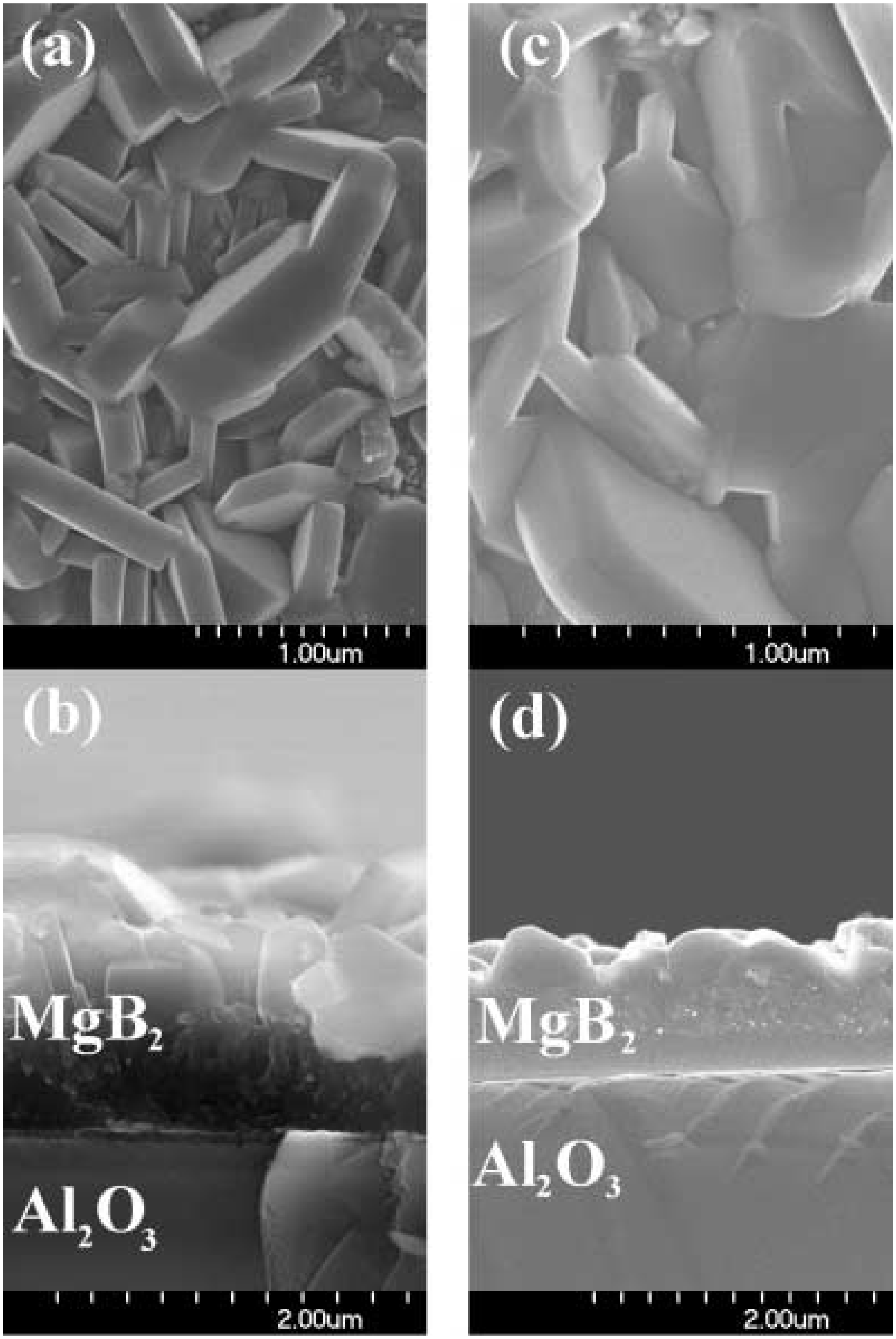

Figure 1

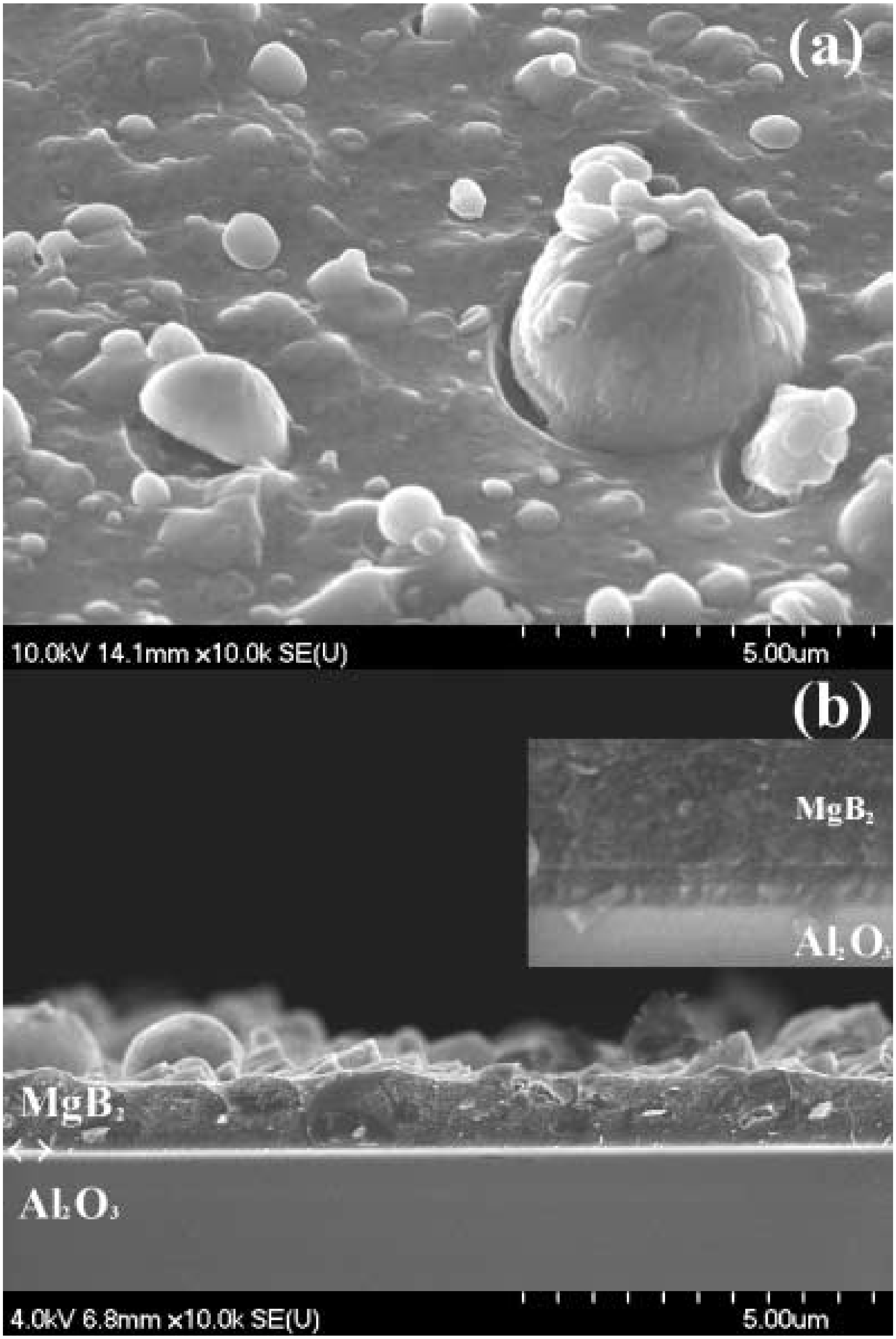

Figure 2

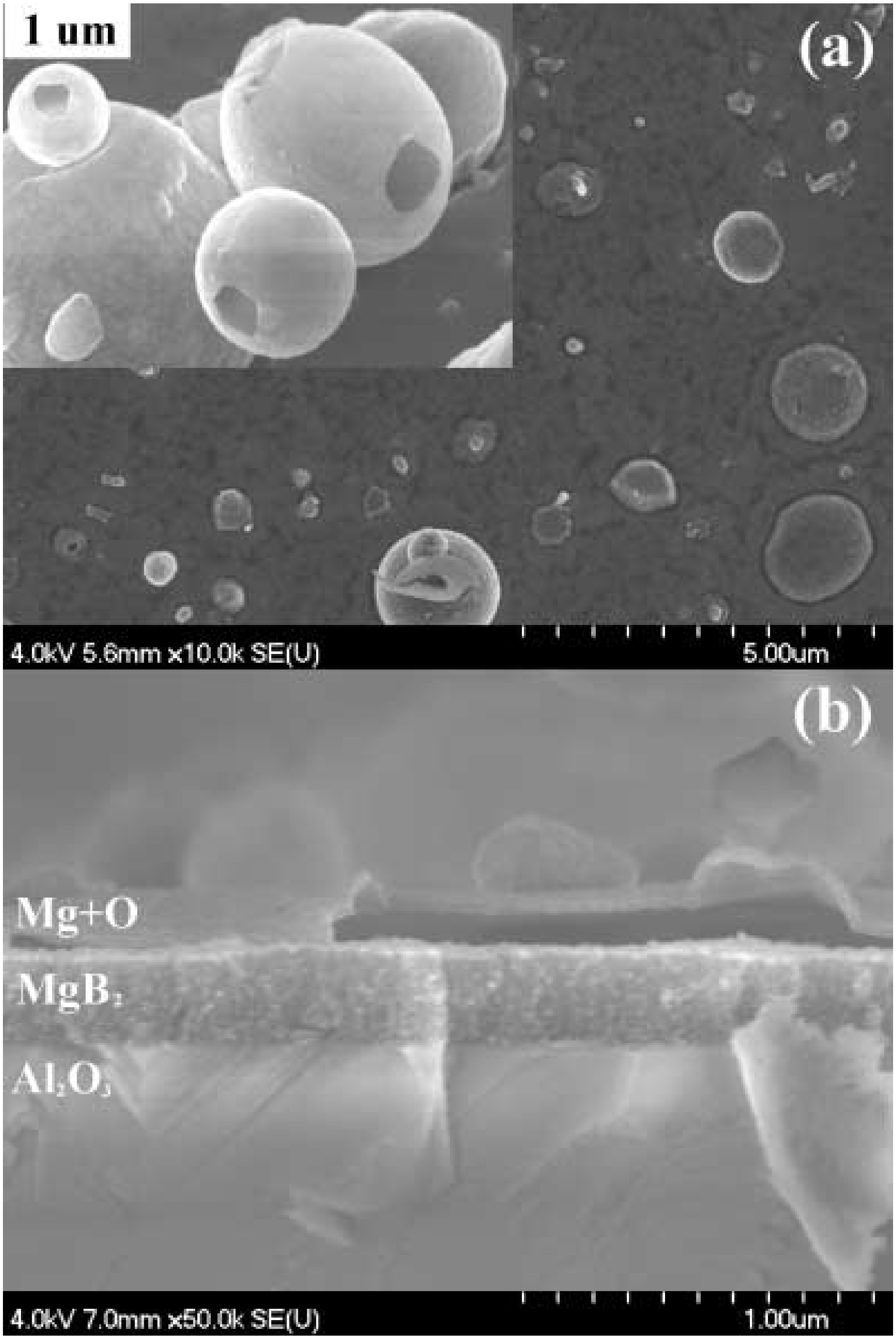

Figure 3

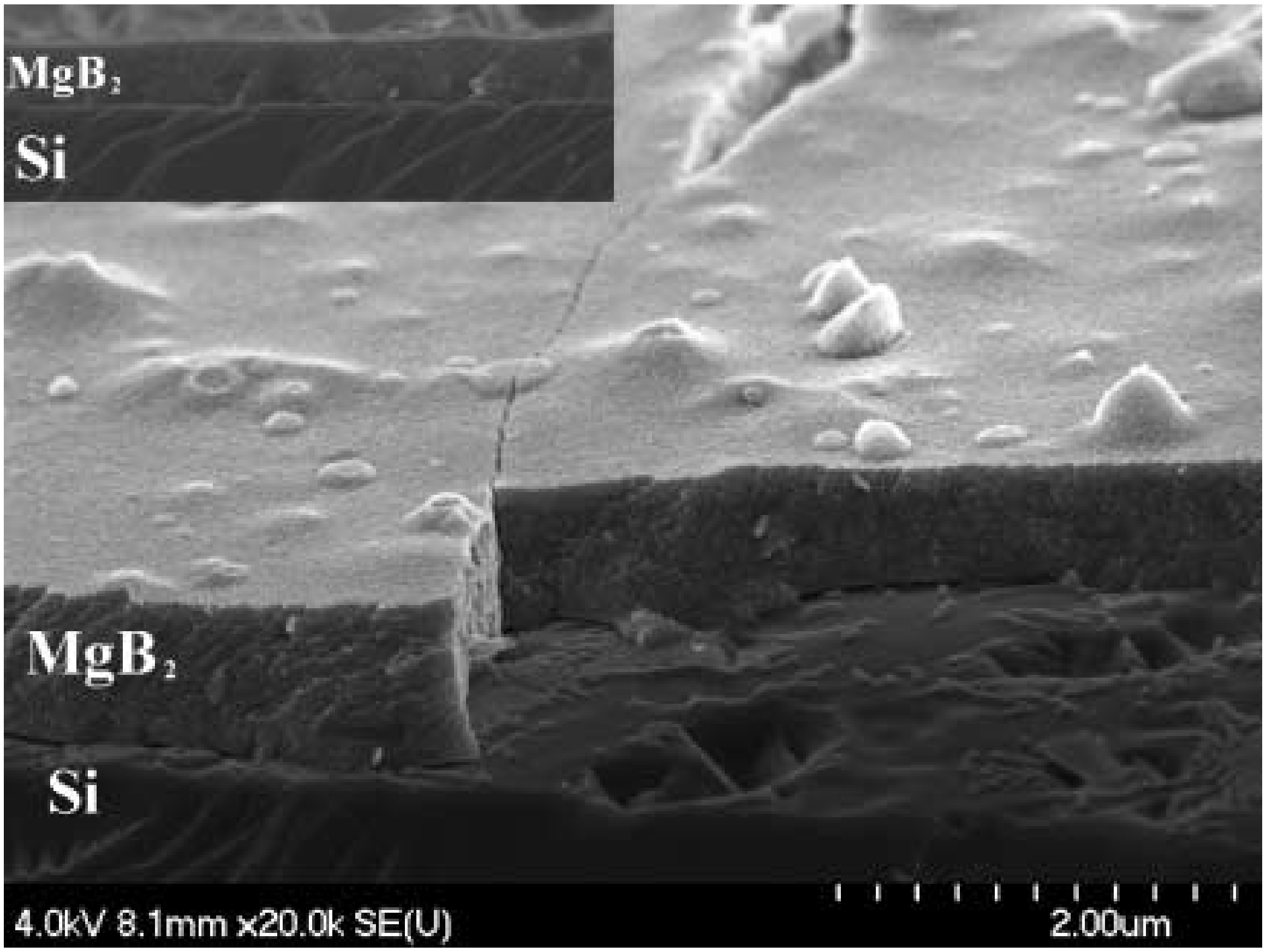

Figure 4